# BSEPACK User's Guide

Meiyue Shao and Chao Yang

Scalable Solvers Group, Computational Research Division
Lawrence Berkeley National Laboratory, Berkeley, California 94720, USA
Email: {myshao,cyang}@lbl.gov

Version 0.1

December 19, 2016

## 1 Introduction

BSEPACK is a parallel ScaLAPACK-style library for solving the Bethe–Salpeter eigenvalue problem

$$\begin{bmatrix} A & B \\ -\overline{B} & -\overline{A} \end{bmatrix} \begin{bmatrix} X_1 & \overline{X_2} \\ X_2 & \overline{X_1} \end{bmatrix} = \begin{bmatrix} X_1 & \overline{X_2} \\ X_2 & \overline{X_1} \end{bmatrix} \begin{bmatrix} \Lambda & 0 \\ 0 & -\Lambda \end{bmatrix},$$

and computing the absorption spectrum

$$\epsilon(\omega) = \begin{bmatrix} d \\ -\overline{d} \end{bmatrix}^* \delta\left(\omega I - \begin{bmatrix} A & B \\ -\overline{B} & -\overline{A} \end{bmatrix}\right) \begin{bmatrix} d \\ -\overline{d} \end{bmatrix},$$

where $A$, $B$, $X_1$, $X_2 \in \mathbb{C}^{2n \times 2n}$, $\Lambda \in \mathbb{R}^{n \times n}$, $d \in \mathbb{C}^n$, and $\delta(\cdot)$ is the Dirac $\delta$-function. The matrix

$$\Omega = \begin{bmatrix} A & B \\ B & A \end{bmatrix}$$

is required to be Hermitian and positive definite.

The library BSEPACK is written in Fortran 90 with MPI, and targets distributed memory HPC systems. This document concerns the usage of BSEPACK. For the description of the algorithm and implementation, we refer to [3, 5, 7].

## 2 Installation

In the following, we provide an installation guide for Unix-like systems.

### 2.1 Prerequisites

To build the library, the following software is required.

- A Fortran 90/95 compiler.

- The MPI library, e.g., OpenMPI or MPICH.
- An optimized BLAS library, e.g., ATLAS or OpenBLAS.
- The LAPACK library, version $\geq 3.4.0$.
- The ScaLAPACK library (including BLACS and PBLAS), version $\geq 2.0.2$.

## 2.2 How to compile the library

**Download location.**
The software can be downloaded from the BSEPACK homepage [4].

**Files in the tar-ball.**
By unpacking the tar-ball through

```
tar xzfv bsepack.tar.gz
```

a directory `BSEPACK/`, which is the root directory of the library, is created with the following files and subdirectories.

```
BSEPACK_UG.pdf   EXAMPLES    Legal.txt   License.txt   MAKE_INC   Makefile
make.inc         README.md   SRC         SSEIG         TESTING
```

Below is an overview of these items.

- `BSEPACK_UG.pdf` The User's Guide of BSEPACK (i.e., this document).
- `EXAMPLES/` This directory contains two simple drivers.
- `Legal.txt` Copyright notice of BSEPACK.
- `License.txt` License agreement of BSEPACK.
- `MAKE_INC/` This directory contains templates of `make.inc`.
- `Makefile` The Makefile for building the library. This file does *not* need to be modified.
- `make.inc` *This is the only file which requires modifications when building the library.* It contains settings for compilers and external libraries used by the Makefile. The user needs to modify this file according to the target computational environment before compiling the library. Templates of this file are provided in the directory `MAKE_INC`.
- `README.md` A shorter version of this document contains a quick installation guide.
- `SRC/` This directory contains source code for all computational routines of the library.
- `SSEIG/` This directory contains a set of BLAS/LAPACK-like subroutines for real skew-symmetric matrices.
- `TESTING/` This directory contains testing examples.

|   |   |   |   |   |   |   |
|---|---|---|---|---|---|---|
| 0 | 1 | 0 | 1 | 0 | 1 | 0 |
| 2 | 3 | 2 | 3 | 2 | 3 | 2 |
| 0 | 1 | 0 | 1 | 0 | 1 | 0 |
| 2 | 3 | 2 | 3 | 2 | 3 | 2 |
| 0 | 1 | 0 | 1 | 0 | 1 | 0 |
| 2 | 3 | 2 | 3 | 2 | 3 | 2 |
| 0 | 1 | 0 | 1 | 0 | 1 | 0 |

Figure 1: The 2D block-cyclic data layout across a $2 \times 2$ processor grid. For example, processor $(0, 1)$ owns all highlighted blocks.

**Build the library.**
After `make.inc` has been properly modified according to the computational environment, the library can be built by

```
make all
```

from the root directory of BSEPACK. This command generates the two library archives `libbsepack.a` and `libsseig.a` in `lib/`, four examples in `EXAMPLES/`, and test programs in `TESTING/`. The script `runall.sh` in `TESTING`, which performs a set tests, needs to be run after the compilation. You may need to modify the MPI execution command in this script, as well as in `SSEIG/runpar.sh`, according to your system (e.g., `mpirun`, `mpiexec`, etc.). Hopefully something similar to the following will be displayed on the screen:

```
% 4096 out of 4096 tests passed!
% 4096 out of 4096 tests passed!
% 8192 out of 8192 tests passed!
% 8192 out of 8192 tests passed!
  ⋮
```

The result is also collected in the file `summary.txt`. If all runs passed the test, then the compilation has been successful.

## 3 Using the package

### 3.1 ScaLAPACK's 2D block-cyclic data layout convention

In ScaLAPACK, the $p = p_r p_c$ processors are usually arranged into a $p_r \times p_c$ grid. Matrices are distributed across the rectangular processor grid in a *2D block-cyclic layout* with block size $m_b \times n_b$ (see Figure 1 for an example). The information regarding the data layout is stored in an *array descriptor* to establish the mapping between the entries of the global matrix and their corresponding locations in the memory hierarchy. We adopt ScaLAPACK's data layout convention in BSEPACK. In addition we require that the $n \times n$ input matrices $A$ and $B$ have identical data layout with square data blocks (i.e., $m_b = n_b$), and the output matrix $X$ has the same block factor as that of $A$ or $B$. The processor grid, however, does *not* need to be square.

A distributed matrix, $A$, is referenced by two arrays `A` (local matrix entries) and `DESCA` (array descriptor). A typical setting of `DESCA` is listed below.

- `DESCA(1)`: Type of the matrix. In our case, `DESCA(1) = 1` since $A$ is stored as a dense matrix.

- `DESCA(2)`: The handle of the BLACS context.

- `DESCA(3)`, `DESCA(4)`: The size of $A$, i.e., `DESCA(3) = DESCA(4) = ` $n$.

- `DESCA(5)`, `DESCA(6)`: Blocking factors $m_b$ and $n_b$. We require that `DESCA(5) = DESCA(6)`.

- `DESCA(7)`, `DESCA(8)`: The process row and column that contain $h_{11}$. Usually, `DESCA(7) = DESCA(8)` $= 0$.

- `DESCA(9)`: Leading dimension of the local part of $A$ on the current processor. This value needs to be at least one, even if the local part is empty.

## 3.2 Computing the spectral decomposition

One functionality of this package is to compute the spectral decomposition of a definite Bethe–Salpeter Hamiltonian matrix

$$H = \begin{bmatrix} A & B \\ -\overline{B} & -\overline{A} \end{bmatrix}. \tag{1}$$

The spectral decomposition is of the form

$$H \begin{bmatrix} X_1 & \overline{X}_2 \\ X_2 & \overline{X}_1 \end{bmatrix} = \begin{bmatrix} X_1 & \overline{X}_2 \\ X_2 & \overline{X}_1 \end{bmatrix} \begin{bmatrix} \Lambda & 0 \\ 0 & -\Lambda \end{bmatrix}, \tag{2}$$

where the eigenvectors are normalized to satisfy

$$\begin{bmatrix} X_1 & -\overline{X}_2 \\ -X_2 & \overline{X}_1 \end{bmatrix}^* \begin{bmatrix} X_1 & \overline{X}_2 \\ X_2 & \overline{X}_1 \end{bmatrix} = I. \tag{3}$$

This can be done by calling the subroutine `PDBSEIG` for a real $H$, and `PZBSEIG` for a complex $H$. The interface of `PDBSEIG`/`PZBSEIG` displayed below follows the convention of LAPACK/ScaLAPACK subroutines [1, 2].

```
      SUBROUTINE PDBSEIG( SOLVER, N, A, IA, JA, DESCA, B, IB, JB, DESCB,
     $                    LAMBDA, X, IX, JX, DESCX, WORK, LWORK, IWORK,
     $                    LIWORK, INFO )
*
*     .. Scalar Arguments ..
      INTEGER           SOLVER, N, IA, JA, IB, JB, IX, JX, LWORK,
     $                  LIWORK, INFO
*     ..
*     .. Array Arguments ..
      DOUBLE PRECISION  A( * ), B( * ), LAMBDA( * ), X( * ), WORK( * )
      INTEGER           DESCA( * ), DESCB( * ), DESCX( * ), IWORK( * )
```

```
      SUBROUTINE PZBSEIG( SOLVER, N, A, IA, JA, DESCA, B, IB, JB, DESCB,
     $                    LAMBDA, X, IX, JX, DESCX, WORK, LWORK, RWORK,
     $                    LRWORK, IWORK, LIWORK, INFO )
*
*     .. Scalar Arguments ..
      INTEGER           SOLVER, N, IA, JA, IB, JB, IX, JX, LWORK,
     $                  LRWORK, LIWORK, INFO
*     ..
*     .. Array Arguments ..
      COMPLEX*16        A( * ), B( * ), X( * ), WORK( * )
      DOUBLE PRECISION  LAMBDA( * ), RWORK( * )
      INTEGER           DESCA( * ), DESCB( * ), DESCX( * ), IWORK( * )
```

The interfaces of `PDBSEIG` and `PZBSEIG` are nearly identical, except that `PZBSEIG` requires one extra workspace `RWORK` of length `LRWORK`. Examples of calling these subroutines are provided in `EXAMPLES/eigenvalue_real.f` and `EXAMPLES/eigenvalue_complex.f`. Similar to most LAPACK/ScaLAPACK subroutines, we advice that `PDBSEIG`/`PZBSEIG` is called twice—the first call for performing a workspace query (by setting `LWORK = −1`) and the second call for actual computation.

**A detailed list of the arguments for `PDBSEIG`.**

- `SOLVER`: (global input) `INTEGER`.
  See Table 2 in Section 3.4 for a full list of supported solvers.

- `N`: (global input) `INTEGER`.
  The order of $A$ (and $B$).

- `A`: (local input) `DOUBLE PRECISION` array of dimension (`DESCA(9)`,*).
  `IA,JA`: (global input) `INTEGER`.
  `DESCA`: (global and local input) `INTEGER` array descriptor of dimension 9.
  `A`, `IA`, `JA`, and `DESCA` define the distributed matrix $A$.
  On entry, the lower triangular part of `A` contains that of the real symmetric matrix $A$, and its strictly upper triangular part is not referenced.
  If Tamm–Dancoff approximation (TDA) is used, `A` is destroyed on exit.

- B: (local input) DOUBLE PRECISION array of dimension (DESCB(9),*).
  IB,JB: (global input) INTEGER.
  DESCB: (global and local input) INTEGER array descriptor of dimension 9.
  B, IB, JB, and DESCB define the distributed matrix $B$.
  On entry, the lower triangular part of B contains that of the real symmetric matrix $B$, and its strictly upper triangular part is not referenced.
  If Tamm–Dancoff approximation (TDA) is used, B is not referenced. However, DESCB still needs to be consistent with DESCA.

- LAMBDA: (global output) DOUBLE PRECISION array of dimension N.
  The positive eigenvalues of $H$ defined by $A$ and $B$ as in (1). The eigenvalues are sorted in ascending order.

- X: (local output) DOUBLE PRECISION array of dimension (DESCX(9),*).
  IX,JX: (global input) INTEGER.
  DESCX: (global and local input) INTEGER array descriptor of dimension 9.
  X, IX, JX, and DESCX define the distributed matrix $X$. In current release, only $IX = JX = 1$ is supported.
  On exit, X contains the normalized eigenvectors associated with the positive eigenvalues of $H$, i.e.
  $$X = \begin{bmatrix} X_1 \\ X_2 \end{bmatrix}$$
  as in (2) and satisfies (3). In the case of Tamm–Dancoff approximation (TDA), the $X_2$ block is not referenced.

- WORK: (local workspace) DOUBLE PRECISION array of dimension LWORK.
  LWORK: (local input) INTEGER.
  In case $LWORK = -1$, a workspace query will be performed and on exit, WORK(1) is set to the required length of the double precision workspace. No computation is performed in this case.

- IWORK: (local workspace) INTEGER array of dimension LIWORK.
  LIWORK: (local input) INTEGER.
  In case $LIWORK = -1$, a workspace query will be performed and on exit, IWORK(1) is set to the required length of the integer workspace. No computation is performed in this case.

- INFO: (global output) INTEGER.
  If $INFO = 0$, PDBSEIG returns successfully.
  If $INFO < 0$, let $i = -INFO$, then the $i$-th argument had an illegal value.
  If $INFO > 0$, the eigensolver failed to converge. (This is a rare case.)

**A detailed list of the arguments for PZBSEIG.**

- SOLVER: (global input) INTEGER.
  See Table 3 in Section 3.4 for a full list of supported solvers.

- N: (global input) INTEGER.
  The order of $A$ (and $B$).

- A: (local input) COMPLEX*16 array of dimension (DESCA(9),*).
  IA,JA: (global input) INTEGER.
  DESCA: (global and local input) INTEGER array descriptor of dimension 9.
  A, IA, JA, and DESCA define the distributed matrix $A$.
  On entry, the lower triangular part of A contains that of the Hermitian matrix $A$, and its strictly upper triangular part is not referenced.
  If Tamm–Dancoff approximation (TDA) is used, A is destroyed on exit.

- B: (local input) COMPLEX*16 array of dimension (DESCB(9),*).
  IB,JB: (global input) INTEGER.
  DESCB: (global and local input) INTEGER array descriptor of dimension 9.
  B, IB, JB, and DESCB define the distributed matrix $B$.
  On entry, the lower triangular part of B contains that of the complex symmetric matrix $B$, and its strictly upper triangular part is not referenced.
  If Tamm–Dancoff approximation (TDA) is used, B is not referenced. However, DESCB still needs to be consistent with DESCA.

- LAMBDA: (global output) DOUBLE PRECISION array of dimension N.
  The positive eigenvalues of $H$ defined by $A$ and $B$ as in (1). The eigenvalues are sorted in ascending order.

- X: (local output) COMPLEX*16 array of dimension (DESCX(9),*).
  IX,JX: (global input) INTEGER.
  DESCX: (global and local input) INTEGER array descriptor of dimension 9.
  X, IX, JX, and DESCX define the distributed matrix $X$. In current release, only IX = JX = 1 is supported.
  On exit, X contains the normalized eigenvectors associated with the positive eigenvalues of $H$, i.e.
  $$X = \begin{bmatrix} X_1 \\ X_2 \end{bmatrix}$$
  as in (2) and satisfies (3). In the case of Tamm–Dancoff approximation (TDA), the $X_2$ block is not referenced.

- WORK: (local workspace) COMPLEX*16 array of dimension LWORK.
  LWORK: (local input) INTEGER.
  In case LWORK $= -1$, a workspace query will be performed and on exit, WORK(1) is set to the required length of the double precision complex workspace. No computation is performed in this case.

- RWORK: (local workspace) DOUBLE PRECISION array of dimension LRWORK.
  LRWORK: (local input) INTEGER.
  In case LRWORK $= -1$, a workspace query will be performed and on exit, RWORK(1) is set to the required length of the double precision real workspace. No computation is performed in this case.

- IWORK: (local workspace) INTEGER array of dimension LIWORK.
  LIWORK: (local input) INTEGER.
  In case LIWORK $= -1$, a workspace query will be performed and on exit, IWORK(1) is set to the required length of the integer workspace. No computation is performed in this case.

- `INFO`: (global output) `INTEGER`.
  If `INFO` = 0, `PZBSEIG` returns successfully.
  If `INFO` < 0, let $i = -\texttt{INFO}$, then the $i$-th argument had an illegal value.
  If `INFO` > 0, the eigensolver failed to converge. (This is a rare case.)

## 3.3 Computing the absorption spectrum

Another functionality of this package is to compute the absorption spectrum

$$\epsilon(\omega) = \begin{bmatrix} d \\ -\bar{d} \end{bmatrix}^* \delta\left(\omega I - \begin{bmatrix} A & B \\ -B & -\bar{A} \end{bmatrix}\right) \begin{bmatrix} d \\ \bar{d} \end{bmatrix}$$

over a number of sampling points $\omega$. This can be done by calling the subroutine `PDBSABSP` for a real $H$ and $d$, and `PZBSABSP` for a complex $H$ and $d$. Two options are provided: one is to compute $\epsilon(\sigma)$ by diagonalizing $H$, and the other is to estimate $\epsilon(\sigma)$ by Lanczos algorithm. Note that the former is much more expensive compared to the latter, though more accurate also.

In practice the Dirac $\delta$-function must be regularized. BSEPACK supports two types of broadening of the $\delta$-function as follows, with broadening factor $\sigma > 0$.

1. Gaussian function
$$f_\sigma(x) = \frac{1}{\sqrt{2\pi}\,\sigma} \exp\left(-\frac{x^2}{2\sigma^2}\right); \tag{4}$$

2. Lorentzian function
$$f_\sigma(x) = \frac{\sigma}{\pi(x^2 + \sigma^2)}. \tag{5}$$

Hence, a broadened absorption spectrum

$$\epsilon_\sigma(\omega) = \begin{bmatrix} d \\ -\bar{d} \end{bmatrix}^* f_\sigma\left(\omega I - \begin{bmatrix} A & B \\ -B & -\bar{A} \end{bmatrix}\right) \begin{bmatrix} d \\ \bar{d} \end{bmatrix} \tag{6}$$

is actually computed.

The interface of `PDBSABSP`/`PZBSABSP` displayed below also follows the convention of LAPACK/ScaLAPACK. A workspace query call is recommended before the call for actual computation. As a remark, even if the Lanczos algorithm is used, all matrices are treated as dense matrices in the current release.

```
      SUBROUTINE PDBSABSP( SOLVER, N, NPTS, SIGMA, OMEGA, EPS, A, IA,
     $                     JA, DESCA, B, IB, JB, DESCB, LAMBDA, X, IX,
     $                     JX, DESCX, D, ID, JD, DESCD, ALPHA, BETA,
     $                     RESUME, ITMAX, WORK, LWORK, IWORK, LIWORK,
     $                     INFO )
*
      INTEGER            SOLVER, N, NPTS, IA, JA, IB, JB, IX, JX, ID,
     $                   JD, RESUME, ITMAX, LWORK, LIWORK, INFO
      DOUBLE PRECISION   SIGMA
*     ..
*     .. Array Arguments ..
      DOUBLE PRECISION   OMEGA( * ), EPS( * ), LAMBDA( * ), A( * ),
     $                   B( * ), X( * ), D( * ), ALPHA( * ), BETA( * ),
     $                   WORK( * )
      INTEGER            DESCA( * ), DESCB( * ), DESCX( * ), DESCD( * ),
     $                   IWORK( * )
```

```
      SUBROUTINE PZBSABSP( SOLVER, N, NPTS, SIGMA, OMEGA, EPS, A, IA,
     $                     JA, DESCA, B, IB, JB, DESCB, LAMBDA, X, IX,
     $                     JX, DESCX, D, ID, JD, DESCD, ALPHA, BETA,
     $                     RESUME, ITMAX, WORK, LWORK, RWORK, LRWORK,
     $                     IWORK, LIWORK, INFO )
*
*     .. Scalar Arguments ..
      INTEGER            SOLVER, N, NPTS, IA, JA, IB, JB, IX, JX, ID,
     $                   JD, RESUME, ITMAX, LWORK, LRWORK, LIWORK, INFO
      DOUBLE PRECISION   SIGMA
*     ..
*     .. Array Arguments ..
      DOUBLE PRECISION   OMEGA( * ), EPS( * ), LAMBDA( * ), ALPHA( * ),
     $                   BETA( * ), RWORK( * )
      COMPLEX*16         A( * ), B( * ), X( * ), D( * ), WORK( * )
      INTEGER            DESCA( * ), DESCB( * ), DESCX( * ), DESCD( * ),
     $                   IWORK( * )
```

**A detailed list of the arguments for PDBSABSP.**

- SOLVER: (global input) INTEGER.
  See Table 4 in Section 3.4 for a full list of supported solvers.

- N: (global input) INTEGER.
  The order of $A$ (and $B$).

- NPTS: (global input) INTEGER.
  Number of sampling points for $\omega$.

- SIGMA: (global input) DOUBLE PRECISION.
  The broadening factor in the approximation of the $\delta$-function.

- OMEGA: (global input) DOUBLE PRECISION array of dimension NPTS.
  The sampling points for $\omega$.

- EPS: (global output) DOUBLE PRECISION array of dimension NPTS.
  The broadened absorption spectrum $\epsilon_\sigma(\omega)$ evaluated at the sampling points of $\omega$.

- A: (local input) DOUBLE PRECISION array of dimension (DESCA(9),*).
  IA,JA: (global input) INTEGER.
  DESCA: (global and local input) INTEGER array descriptor of dimension 9.
  A, IA, JA, and DESCA define the distributed matrix $A$.
  On entry, the lower triangular part of A contains that of the real symmetric matrix $A$, and its strictly upper triangular part is not referenced.
  If Tamm–Dancoff approximation (TDA) and full diagonalization are both specified, A is destroyed on exit.

- B: (local input) DOUBLE PRECISION array of dimension (DESCB(9),*).
  IB,JB: (global input) INTEGER.
  DESCB: (global and local input) INTEGER array descriptor of dimension 9.
  B, IB, JB, and DESCB define the distributed matrix $B$.
  On entry, the lower triangular part of B contains that of the real symmetric matrix $B$, and its strictly upper triangular part is not referenced.
  If Tamm–Dancoff approximation (TDA) is used, B is not referenced. However, DESCB still needs to be consistent with DESCA.

- LAMBDA: (global output) DOUBLE PRECISION array.
  In case of full diagonalization, LAMBDA, of dimension N, contains the positive eigenvalues of $H$ defined by $A$ and $B$ as in (1). In case of Lanczos algorithm, LAMBDA, of dimension $\min(\mathtt{N}, \mathtt{ITMAX})$, contains the Ritz values. The eigenvalues or Ritz values are sorted in ascending order.

- X: (local output) DOUBLE PRECISION array of dimension (DESCX(9),*).
  IX,JX: (global input) INTEGER.
  DESCX: (global and local input) INTEGER array descriptor of dimension 9.
  X, IX, JX, and DESCX define the distributed matrix $X$. In current release, only IX = JX = 1 is supported.
  On exit, X contains the normalized eigenvectors associated with the positive eigenvalues of $H$ (for full diagonalization) or Lanczos vectors (for Lanczos algorithm). In the case of Tamm–Dancoff approximation (TDA), the $X_2$ block is not referenced.

- D: (local input) DOUBLE PRECISION array of dimension (DESCD(9),*).
  ID,JD: (global input) INTEGER.
  DESCD: (global and local input) INTEGER array descriptor of dimension 9.
  D, ID, JD, and DESCD define the distributed matrix $D$. D has only one column of length $n$, and stores the optical transition vector $d$.

- ALPHA: (global output) DOUBLE PRECISION array of dimension ITMAX.
  On exit, ALPHA contains the diagonal entries of the tridiagonal matrix produced by the Lanczos procedure. ALPHA is not referenced in case of full diagonalization.

- BETA: (global output) DOUBLE PRECISION array of dimension ITMAX.
  On exit, BETA contains the subdiagonal entries of the tridiagonal matrix produced by the Lanczos procedure. BETA is not referenced in case of full diagonalization.

- RESUME: (global input) INTEGER.
  This argument is not supported in the current release.

- ITMAX: (global input) INTEGER.
  The maximum number of Lanczos iterations. ITMAX is not referenced in case of full diagonalization.

- WORK: (local workspace) DOUBLE PRECISION array of dimension LWORK.
  LWORK: (local input) INTEGER.
  In case LWORK $= -1$, a workspace query will be performed and on exit, WORK(1) is set to the required length of the double precision workspace. No computation is performed in this case.

- IWORK: (local workspace) INTEGER array of dimension LIWORK.
  LIWORK: (local input) INTEGER.
  In case LIWORK $= -1$, a workspace query will be performed and on exit, IWORK(1) is set to the required length of the integer workspace. No computation is performed in this case.

- INFO: (global output) INTEGER.
  If INFO $= 0$, PDBSABSP returns successfully.
  If INFO $< 0$, let $i = -$INFO, then the $i$-th argument had an illegal value.
  If INFO $> 0$, the eigensolver PDBSEIG failed to converge (for full diagonalization) or the Lanczos process breaks down after INFO steps and returns successfully (for Lanczos algorithm).

**A detailed list of the arguments for PZBSABSP.**

- SOLVER: (global input) INTEGER.
  See Table 5 in Section 3.4 for a full list of supported solvers.

- N: (global input) INTEGER.
  The order of $A$ (and $B$).

- NPTS: (global input) INTEGER.
  Number of sampling points for $\omega$.

- SIGMA: (global input) DOUBLE PRECISION.
  The broadening factor in the approximation of the $\delta$-function.

- OMEGA: (global input) DOUBLE PRECISION array of dimension NPTS.
  The sampling points for $\omega$.

- EPS: (global output) DOUBLE PRECISION array of dimension NPTS.
  The broadened absorption spectrum $\epsilon_\sigma(\omega)$ evaluated at the sampling points of $\omega$.

- A: (local input) COMPLEX*16 array of dimension (DESCA(9),*).
  IA,JA: (global input) INTEGER.
  DESCA: (global and local input) INTEGER array descriptor of dimension 9.
  A, IA, JA, and DESCA define the distributed matrix $A$.
  On entry, the lower triangular part of A contains that of the Hermitian matrix $A$, and its strictly upper triangular part is not referenced.
  If Tamm–Dancoff approximation (TDA) and full diagonalization are both specified, A is destroyed on exit.

- B: (local input) COMPLEX*16 array of dimension (DESCB(9),*).
  IB,JB: (global input) INTEGER.
  DESCB: (global and local input) INTEGER array descriptor of dimension 9.
  B, IB, JB, and DESCB define the distributed matrix $B$.
  On entry, the lower triangular part of B contains that of the complex symmetric matrix $B$, and its strictly upper triangular part is not referenced.
  If Tamm–Dancoff approximation (TDA) is used, B is not referenced. However, DESCB still needs to be consistent with DESCA.

- LAMBDA: (global output) DOUBLE PRECISION array.
  In case of full diagonalization, LAMBDA, of dimension N, contains the positive eigenvalues of $H$ defined by $A$ and $B$ as in (1). In case of Lanczos algorithm, LAMBDA, of dimension $\min(\text{N}, \text{ITMAX})$, contains the Ritz values. The eigenvalues or Ritz values are sorted in ascending order.

- X: (local output) COMPLEX*16 array of dimension (DESCX(9),*).
  IX,JX: (global input) INTEGER.
  DESCX: (global and local input) INTEGER array descriptor of dimension 9.
  X, IX, JX, and DESCX define the distributed matrix $X$. In current release, only IX = JX = 1 is supported.
  On exit, X contains the normalized eigenvectors associated with the positive eigenvalues of $H$ (for full diagonalization) or Lanczos vectors (for Lanczos algorithm). In the case of Tamm–Dancoff approximation (TDA), the $X_2$ block is not referenced.

- D: (local input) COMPLEX*16 array of dimension (DESCD(9),*).
  ID,JD: (global input) INTEGER.
  DESCD: (global and local input) INTEGER array descriptor of dimension 9.
  D, ID, JD, and DESCD define the distributed matrix $D$. D has only one column of length $n$, and stores the optical transition vector $d$.

- ALPHA: (global output) DOUBLE PRECISION array of dimension ITMAX.
  On exit, ALPHA contains the diagonal entries of the tridiagonal matrix produced by the Lanczos procedure. ALPHA is not referenced in case of full diagonalization.

- BETA: (global output) DOUBLE PRECISION array of dimension ITMAX.
  On exit, BETA contains the subdiagonal entries of the tridiagonal matrix produced by the Lanczos procedure. BETA is not referenced in case of full diagonalization.

- RESUME: (global input) INTEGER.
  This argument is not supported in the current release.

- ITMAX: (global input) INTEGER.
  The maximum number of Lanczos iterations. ITMAX is not referenced in case of full diagonalization.

- WORK: (local workspace) COMPLEX*16 array of dimension LWORK.
  LWORK: (local input) INTEGER.
  In case LWORK $= -1$, a workspace query will be performed and on exit, WORK(1) is set to the required length of the double precision complex workspace. No computation is performed in this case.

- RWORK: (local workspace) DOUBLE PRECISION array of dimension LRWORK.
  LRWORK: (local input) INTEGER.
  In case LRWORK $= -1$, a workspace query will be performed and on exit, RWORK(1) is set to the required length of the double precision real workspace. No computation is performed in this case.

- IWORK: (local workspace) INTEGER array of dimension LIWORK.
  LIWORK: (local input) INTEGER.
  In case LIWORK $= -1$, a workspace query will be performed and on exit, IWORK(1) is set to the required length of the integer workspace. No computation is performed in this case.

- INFO: (global output) INTEGER.
  If INFO $= 0$, PZBSABSP returns successfully.
  If INFO $< 0$, let $i = -$INFO, then the $i$-th argument had an illegal value.
  If INFO $> 0$, the eigensolver PZBSEIG failed to converge (for full diagonalization) or the Lanczos process breaks down after INFO steps and returns successfully (for Lanczos algorithm).

## 3.4 Lists of supported solvers

In the computational subroutines, the first argument, SOLVER, is an integer that specifies the choice of algorithm or algorithmic variant for the computation. In general, there are four classes of options, and SOLVER is the combination of up to four options, one from each class.[*] The options are defined in the file solver.f, which needs to be included whenever calling the computational subroutines. A list of options is shown in Table 1. For example, when calling PZBSABSP,

$$\text{SOLVER} = \text{BSE\_FULLBSE} + \text{BSE\_LANCZOS} + \text{BSE\_LORENTIAN} + \text{BSE\_QUADAVGGAUSS}$$

means using the Lanczos algorithm with generalized averaged Gauss quadrature to estimate the absorption spectrum broadened by the Lorentzian for complex full BSE. Some options can be omitted, especially when they do not make sense. For instance, neither BSE_GAUSSIAN nor BSE_LORENTZIAN should be specified when calling PDBSEIG. Lists of all supported solvers are provided in Tables 2–5.

---

[*]Future releases may allow multiple options from Class 4.
[†]There are known bugs with this option due to bugs in PZHEEV.
[‡]There are known bugs with this option due to bugs in PZHEEV.
[§]There are known bugs with this option due to bugs in PZHEEV.

Table 1: List of options for the argument `SOLVER`.

| Name | Class | Meaning |
|---|---|---|
| `BSE_FULLBSE` | 1 | Full BSE solver |
| `BSE_TDA` | 1 | Tamm–Dancoff approximation (i.e., set $B=0$) |
| `BSE_DIRECT` | 2 | Diagonalize the BSE Hamiltonian |
| `BSE_LANCZOS` | 2 | Lanczos algorithm (for estimating the absorption spectrum) |
| `BSE_GAUSSIAN` | 3 | Use the Gaussian function (4) to approximate the $\delta$-function |
| `BSE_LORENTZIAN` | 3 | Use the Lorentzian function (5) to approximate the $\delta$-function |
| `BSE_PRODUCT` | 4 | Solve the real BSE through the product form (see [5]) |
| `BSE_SVD` | 4 | Solve the real BSE using SVD (see [5]) |
| `BSE_LAPACK_HEEV` | 4 | Use `(P)DSYEV`/`(P)ZHEEV` to diagonalize the matrix (see [1, 2]) |
| `BSE_LAPACK_HEEVD` | 4 | Use `(P)DSYEVD`/`(P)ZHEEVD` to diagonalize the matrix (see [1, 2]) |
| `BSE_LAPACK_HEEVR` | 4 | Use `(P)DSYEVR`/`(P)ZHEEVR` to diagonalize the matrix (see [1, 2]) |
| `BSE_LAPACK_HEEVX` | 4 | Use `(P)DSYEVX`/`(P)ZHEEVX` to diagonalize the matrix (see [1, 2]) |
| `BSE_QUADAVGGAUSS` | 4 | Use generalized averaged Gauss quadrature to estimate (6) (see [7]) |

Table 2: List of supported solvers for `PDBSEIG`.

| | |
|---|---|
| Full BSE | `BSE_FULLBSE+BSE_DIRECT` |
| | `BSE_FULLBSE+BSE_DIRECT+BSE_PRODUCT` |
| | `BSE_FULLBSE+BSE_DIRECT+BSE_SVD` |
| TDA | `BSE_TDA+BSE_DIRECT` |
| | `BSE_TDA+BSE_DIRECT+BSE_LAPACK_HEEV` |
| | `BSE_TDA+BSE_DIRECT+BSE_LAPACK_HEEVD` |
| | `BSE_TDA+BSE_DIRECT+BSE_LAPACK_HEEVR` |
| | `BSE_TDA+BSE_DIRECT+BSE_LAPACK_HEEVX` |

Table 3: List of supported solvers for `PZBSEIG`.

| | |
|---|---|
| Full BSE | `BSE_FULLBSE+BSE_DIRECT` |
| TDA | `BSE_TDA+BSE_DIRECT` |
| | `BSE_TDA+BSE_DIRECT+BSE_LAPACK_HEEV`[†] |
| | `BSE_TDA+BSE_DIRECT+BSE_LAPACK_HEEVD` |
| | `BSE_TDA+BSE_DIRECT+BSE_LAPACK_HEEVR` |
| | `BSE_TDA+BSE_DIRECT+BSE_LAPACK_HEEVX` |

## 3.5 Example programs

We provide four simple examples in the directory `EXAMPLES/`, two for real matrices and two for complex matrices.

The program `absorption_real.f` reads two $128 \times 128$ $A$ and $B$, the optical transition vector $d$, and the broadening factor $\sigma$ from the input file `input_real.txt`, and estimates the (broadened) absorption spectrum (6) on 1024 sampling points of $\omega$ by the Lanczos algorithm. The format of the input file is described in Table 6. The program `eigenvalue_real.f` reads $A$ and $B$ from `input_real.txt` (the rest of the input file is discarded), and computes the spectral decomposi-tion (2).

Table 4: List of supported solvers for `PDBSABSP`.

| | |
|---|---|
| Full BSE/diagonalization | `BSE_FULLBSE+BSE_DIRECT+BSE_GAUSSIAN` |
| | `BSE_FULLBSE+BSE_DIRECT+BSE_GAUSSIAN+BSE_PRODUCT` |
| | `BSE_FULLBSE+BSE_DIRECT+BSE_GAUSSIAN+BSE_SVD` |
| | `BSE_FULLBSE+BSE_DIRECT+BSE_LORENTZIAN` |
| | `BSE_FULLBSE+BSE_DIRECT+BSE_LORENTZIAN+BSE_PRODUCT` |
| | `BSE_FULLBSE+BSE_DIRECT+BSE_LORENTZIAN+BSE_SVD` |
| Full BSE/Lanczos | `BSE_FULLBSE+BSE_LANCZOS+BSE_GAUSSIAN` |
| | `BSE_FULLBSE+BSE_LANCZOS+BSE_GAUSSIAN+BSE_QUADAVGGAUSS` |
| | `BSE_FULLBSE+BSE_LANCZOS+BSE_LORENTZIAN` |
| | `BSE_FULLBSE+BSE_LANCZOS+BSE_LORENTZIAN+BSE_QUADAVGGAUSS` |
| TDA/diagonalization | `BSE_TDA+BSE_DIRECT+BSE_GAUSSIAN` |
| | `BSE_TDA+BSE_DIRECT+BSE_LORENTZIAN` |
| | `BSE_TDA+BSE_DIRECT+BSE_GAUSSIAN+BSE_LAPACK_HEEV` |
| | `BSE_TDA+BSE_DIRECT+BSE_LORENTZIAN+BSE_LAPACK_HEEV` |
| | `BSE_TDA+BSE_DIRECT+BSE_GAUSSIAN+BSE_LAPACK_HEEVD` |
| | `BSE_TDA+BSE_DIRECT+BSE_LORENTZIAN+BSE_LAPACK_HEEVD` |
| | `BSE_TDA+BSE_DIRECT+BSE_GAUSSIAN+BSE_LAPACK_HEEVR` |
| | `BSE_TDA+BSE_DIRECT+BSE_LORENTZIAN+BSE_LAPACK_HEEVR` |
| | `BSE_TDA+BSE_DIRECT+BSE_GAUSSIAN+BSE_LAPACK_HEEVX` |
| | `BSE_TDA+BSE_DIRECT+BSE_LORENTZIAN+BSE_LAPACK_HEEVX` |
| TDA/Lanczos | `BSE_TDA+BSE_LANCZOS+BSE_GAUSSIAN` |
| | `BSE_TDA+BSE_LANCZOS+BSE_GAUSSIAN+BSE_QUADAVGGAUSS` |
| | `BSE_TDA+BSE_LANCZOS+BSE_LORENTZIAN` |
| | `BSE_TDA+BSE_LANCZOS+BSE_LORENTZIAN+BSE_QUADAVGGAUSS` |

The programs `absorption_complex.f` and `eigenvalue_complex.f` are similar to their real counterparts `absorption_real.f` and `eigenvalue_real.f`, respectively. The corresponding input file `input_complex.txt` contains complex matrices with $n = 32$.

Note that the sampling points of $\omega$ are not read from the input file. If you wish to handle your own input data format, you will have to modify the example programs.

# 4 Terms of Usage

BSEPACK is released under a modified BSD license; see `License.txt` for details. In addition, any use of the library should be acknowledged by citing the corresponding publication as follows:

- Cite [5, 6] if you use the subroutines `PDBSEIG`/`PZBSEIG`;
- Cite [3, 6, 7] if you use the subroutines `PDBSABSP`/`PZBSABSP`.

Table 5: List of supported solvers for `PZBSABSP`.

| | |
|---|---|
| Full BSE/diagonalization | `BSE_FULLBSE+BSE_DIRECT+BSE_GAUSSIAN` |
| | `BSE_FULLBSE+BSE_DIRECT+BSE_LORENTZIAN` |
| Full BSE/Lanczos | `BSE_FULLBSE+BSE_LANCZOS+BSE_GAUSSIAN` |
| | `BSE_FULLBSE+BSE_LANCZOS+BSE_GAUSSIAN+BSE_QUADAVGGAUSS` |
| | `BSE_FULLBSE+BSE_LANCZOS+BSE_LORENTZIAN` |
| | `BSE_FULLBSE+BSE_LANCZOS+BSE_LORENTZIAN+BSE_QUADAVGGAUSS` |
| TDA/diagonalization | `BSE_TDA+BSE_DIRECT+BSE_GAUSSIAN` |
| | `BSE_TDA+BSE_DIRECT+BSE_LORENTZIAN` |
| | `BSE_TDA+BSE_DIRECT+BSE_GAUSSIAN+BSE_LAPACK_HEEV`[‡] |
| | `BSE_TDA+BSE_DIRECT+BSE_LORENTZIAN+BSE_LAPACK_HEEV`[§] |
| | `BSE_TDA+BSE_DIRECT+BSE_GAUSSIAN+BSE_LAPACK_HEEVD` |
| | `BSE_TDA+BSE_DIRECT+BSE_LORENTZIAN+BSE_LAPACK_HEEVD` |
| | `BSE_TDA+BSE_DIRECT+BSE_GAUSSIAN+BSE_LAPACK_HEEVR` |
| | `BSE_TDA+BSE_DIRECT+BSE_LORENTZIAN+BSE_LAPACK_HEEVR` |
| | `BSE_TDA+BSE_DIRECT+BSE_GAUSSIAN+BSE_LAPACK_HEEVX` |
| | `BSE_TDA+BSE_DIRECT+BSE_LORENTZIAN+BSE_LAPACK_HEEVX` |
| TDA/Lanczos | `BSE_TDA+BSE_LANCZOS+BSE_GAUSSIAN` |
| | `BSE_TDA+BSE_LANCZOS+BSE_GAUSSIAN+BSE_QUADAVGGAUSS` |
| | `BSE_TDA+BSE_LANCZOS+BSE_LORENTZIAN` |
| | `BSE_TDA+BSE_LANCZOS+BSE_LORENTZIAN+BSE_QUADAVGGAUSS` |

Table 6: The format of `input_real.txt` and `input_complex.txt`. For complex $A$, $B$, and $d$, each entry is represented by the real part followed by the imaginary part (in the same line).

| line number(s) | content |
|---|---|
| 1 | two numbers $n$ and $n$ (the dimension of $A$) |
| 2–$(n^2+1)$ | the entries of $A$, one each line |
| $n^2 + 2$ | two numbers $n$ and $n$ (the dimension of $B$) |
| $(n^2+3)$–$(2n^2+2)$ | the entries of $B$, one each line |
| $2n^2 + 3$ | two numbers $n$ and $1$ (the dimension of $d$) |
| $(2n^2+4)$–$(2n^2+n+3)$ | the entries of $d$, one each line |
| $2n^2 + n + 4$ | two numbers $1$ and $1$ (the dimension of $\sigma$) |
| $2n^2 + n + 5$ | the value $\sigma$ (always real even for complex data type) |

# Acknowledgements

The authors thank Fabien Bruneval, Felipe H. da Jornada, Daniel Kressner, Osni Marques, and Hongguo Xu for helpful comments on the implementation.